\documentclass[final,5p,times,twocolumn]{elsarticle}
\pdfoutput=1

\usepackage[utf8]{inputenc}
\usepackage{geometry} 
\usepackage{slashed}
\usepackage{mathrsfs}
\usepackage{amsfonts}
\usepackage{amsmath}
\usepackage{amssymb}
\usepackage{bbm}
\usepackage{verbatim}
\usepackage{graphicx}
\usepackage{mathrsfs}
\usepackage{color}
\usepackage{hyperref}
\usepackage[usenames,dvipsnames]{xcolor}
\usepackage{geometry}
\usepackage{parskip}
\usepackage[normalem]{ulem}
\usepackage{bm}
\usepackage{enumitem}
\usepackage{booktabs} 
\usepackage{array} 
\usepackage{verbatim} 
\usepackage{subfig}

\newcommand{\be}{\begin{equation}}
\newcommand{\ee}{\end{equation}}
\newcommand{\bea}{\begin{eqnarray}}
\newcommand{\eea}{\end{eqnarray}}

\newcommand{\mbf}[1]{\mathbf{#1}}

\newcommand{\trm}[1]{\textrm{#1}}
\newcommand{\avs}{u}
\newcommand{\figref}[1]{Fig. \ref{#1}}

\newcommand{\eqnref}[1]{Eq.\,(\ref{#1})}
\newcommand{\eqnrefs}[2]{Eqs.~(\ref{#1}) and (\ref{#2})}

\newcommand{\tsf}[1]{\textsf{#1}}

\newcommand{\vkap}{\varkappa}
\newcommand{\vphi}{\varphi}

\newcommand{\nn}{\nonumber}

\newcommand{\bi}{\begin{itemize}}
\newcommand{\ei}{\end{itemize}}

\journal{Physics Letters B}

\begin{document}

\begin{frontmatter}
\title{Classical Radiation Reaction in Red-Shifted Harmonics}

\author{B. King}
\address{Centre for Mathematical Sciences, Plymouth University, Plymouth, PL4 8AA, United Kingdom}

\begin{abstract}
The collision of a finite electromagnetic plane wave with an electron subject to the Landau-Lifshitz radiation reaction force is studied. A locally monochromatic approximation is derived and compared to numerical evaluation of the exact plane wave result. Energy and transverse momentum spectra are calculated, which clearly display the red-shifting of harmonic features due to radiation reaction effects. Simple formulas are presented to predict the shifting of harmonic edges, whose position can be used in experiments as evidence of radiation reaction.
\end{abstract}

\end{frontmatter}

\section{Introduction}

In strong-field QED, radiation reaction (RR) can be understood as any chain of processes involving a charge's own EM field, which modifies the momentum spectrum of the scattered charge (see e.g. \cite{Burton:2014wsa,Blackburn:2019rfv,Gonoskov:2021hwf} for reviews). For example, to leading order in the fine structure constant $\alpha$, the total radiation reaction on an electron's expected lightfront momentum  is the sum of single nonlinear Compton scattering plus interference between the (dressed) one-loop self-mass and dressed propagator terms. It was shown in \cite{Krivitsky:1991vt,Higuchi:2002qc,Ilderton:2013dba,Ilderton:2013tb} that the classical limit of the calculation agrees with the classical Landau-Lifshitz equation.
Recent approaches \cite{Torgrimsson:2020gws,Torgrimsson:2021wcj} resum arbitrary-length chains of nonlinear Compton and `propagator $\times$ one-loop' interference links, and have verified  this correspondence to all orders in $\alpha$. One advantage of using a classical approach is that there are equations of motion, such as the Landau-Lifshitz equations, which are physically relevant \cite{Ekman:2021vwg,Ekman:2021eqc} and have a closed-form solution in a plane wave \cite{dipiazza08LL}. Therefore the classical approach can be understood as including a resummation of higher order processes in this limit.

There are two main requirements for the classical limit of RR to be a good approximation to quantum RR for a given set of parameters. First, particle spectra calculated from the classical equations of motions should coincide with those from (strong-field) QED (for reviews, see \cite{dipiazza12,narozhny15,Fedotov:2022ely}). This corresponds to the limit of $\hbar \to 0$. The collision of an electron with a finite plane wave can be described in QED using the (classical, nonlinear) intensity parameter, $\xi$ and the (quantum, linear) energy parameter $\eta$. At high intensities, $\xi \gg 1$, it is often a good approximation to regard the field as `locally constant' \cite{Seipt:2020diz}, and apart from an overall flux factor, the entire probability depends on the strong-field parameter $\chi = \xi \eta$. Therefore, the formal classical limit should correspond to small linear quantum $\eta \ll 1$ and nonlinear quantum $\chi \ll 1$ effects. A second requirement that classical RR be a good approximation can be understood from the correspondence principle, i.e. that the average number of interactions with the charge's own EM field be large. In the classical picture, the charge is continuously radiating, whereas in QED, interaction of the charge's own EM field is stochastic \cite{Neitz:2013qba,yoffe.njp.2015}. This can lead to the situation where the classical RR effect is signficiant, but the average number of emitted photons predicted by QED is small. One outcome is `straggling', where in reality, as predicted by QED, the charge can reach regions of higher external field intensity before emitting than predicted by the classical theory \cite{shen.prl.1972,Blackburn:2014cig}, or if the background EM field is short enough in duration `quenching' where charges pass through the field without emitting at all \cite{Harvey:2016uiy}.

Radiation reaction has been measured at high intensities in collisions of high energy positrons with oriented crystals \cite{Wistisen:2017pgr} and in experiments that collide electron beams with intense laser pulses \cite{cole18,Poder:2017dpw}. It appears as a possible science aim in the planned laser-particle experiment LUXE \cite{Abramowicz:2021zja} and can be searched for in all-optical experiments at the newest generation of high power laser facilities \cite{samarin18,danson19}.

The rate of classical RR in the locally constant field limit has recently appeared in the literature, being used to investigate the Ritus-Narozhny conjecture \cite{Fedotov:2017conjecture,Podszus:2018hnz,Ilderton:2019kqp} in the classical limit \cite{Heinzl:2021mji}, and being written in an angularly-resolved form \cite{Piazza:2021vxi}.

In the current letter, we calculate how outgoing particle spectra from nonlinear Thomson scattering in a finite plane wave are modified by RR. The finite duration of the plane wave generates a divergence that must be regularised \cite{boca09,dinu12}. Classical RR effects have been calculated many times in the literature before; here our interest is to show the position of harmonic structures such as the Thomson edge (the classical equivalent of the `Compton edge' \cite{harvey09}) can be used as an indicator that radiation reaction has taken place. We also derive the locally monochromatic approximation (LMA) \cite{Heinzl:2020ynb} for nonlinear Thomson scattering including classical radiation reaction, to allow for quicker calculation and incorporation into Monte-Carlo numerical simulation codes at moderate intensity parameter, such as Ptarmigan \cite{ptarmiganPaper,ptarmigan}   (in contrast to existing particle-in-cell Monte-Carlo condes at high intensity parameter, see e.g. \cite{hadad.prd.2010,schlegel.njp.2012,thomas.prx.2012,vranic.prl.2014,vranic.cpc.2016}). We will find simple formulas to calculate the red-shifting of lightfront and transverse momentum harmonics due to classical RR.

\section{Radiation spectrum in a finite plane wave using the solution to the Landau-Lifshitz equation}
Scattering calculations in finite plane wave backgrounds, when performed na\"ively, generate divergences due to contributions from pure phase factors. These divergences must be regularised to acquire a physically meaningful result. This is not a feature introduced by quantum effects; we must also regularise in the classical context \cite{dinu12,King:2020hsk}. When radiation reaction is included, the outgoing electron momentum after having propagated through the laser pulse is not equal to the electron's initial momentum. This situation is similar to calculations in plane waves that have a zero-frequency (`DC') component except that here, the change in momentum originates from radiation reaction. In the end, we will arrive at a regularised expression that can be integrated to acquire the outgoing particle spectrum; here and throughout, we will focus on the photon spectrum.

The classical momentum $\ell^{\mu}$ radiated from an electron is:
\be 
\ell^{\mu} = \int \frac{d^{3}k}{(2\pi)^{3}} \frac{|\tilde{\j}(k)|^{2}}{2k^{0}}k^{\mu}, \label{eqn:kmu1}
\ee
where $\tilde{\j}(k)=e\int d\tau\, u(\tau)\exp[ik\cdot x(\tau)]$ is the Fourier-transformed electron current ($e<0$ is the electron charge), $\tau$ is the proper time, $x(\tau)$  is the electron trajectory and $u(\tau)$ is the electron velocity. We specify this to an electron moving in a plane wave, with frequency $\vkap^{0}$, wavevector $\vkap=\vkap^{0}n$ with $n^{2}=0$, phase $\vphi=\vkap\cdot x$ and scaled potential $eA(\vphi)=a(\vphi)$ with $\vkap\cdot a=0$. We also define the intensity parameter $\xi$ through $a=m\xi \Psi(\vphi)$, where $|\Psi^{\mu}(\vphi)|\leq 1$. Without loss of generality, we assume the plane wave has finite support\footnote{Whilst this would imply no DC components in the background field, as mentioned in the text, there is still a net change in electron momentum due to RR.} so that $\Psi(\vphi)=0$ when  $\vphi<0$ or $\vphi>\Phi$, meaning $\Phi$ defines the phase duration of the plane wave. Then if the electron  experiences radiation reaction described by the Landau Lifshitz equation, to calculate the current, we invoke the solution \cite{dipiazza08LL} (see also \cite{Heintzmann:1972mn}) for the velocity:
\be 
u = \frac{1}{\mathcal{R}} \left[u_{0}-\mathcal{I}+\frac{n}{2\,n\cdot u_{0}}\left(2\,\mathcal{I}\cdot u_{0} - \mathcal{I}^{2}+\mathcal{R}^{2}-1\right)\right], \label{eqn:ub}
\ee
where $u_{0}$ is the initial velocity before the electron has scattered with the plane wave and the phase-dependent functions are defined:
\bea
\mathcal{R} = 1+\nu X(\vphi); \qquad X(\vphi) = \frac{1}{\Phi}\int^{\vphi}\,\Psi'(\phi)\cdot\Psi'(\phi)~d\phi, \label{eqn:R1}
\eea
quantifying the relative change in the lightfront component ${u_{0}\cdot n/u \cdot n}$, and:
\bea 
\hspace{-0.5cm}\mathcal{I} = \frac{a(\vphi)}{m} + \nu Y(\vphi); \quad Y(\vphi) = \frac{\Psi'(\vphi)}{\xi\Phi} + \xi\Phi \int^{\vphi} X(\phi)\frac{\Psi'(\phi)}{\Phi}d\phi,\nn \\\hspace{-1cm}\phantom{.}\hspace{-1cm} \label{eqn:Ic1}
\eea
quantifying the change in the transverse velocity components. In the above, we have defined a classical parameter quantifying radiation reaction:
\bea 
\nu = \frac{2}{3}\alpha\eta\xi^{2}\Phi
\eea  
such that the integrals in \eqnrefs{eqn:R1}{eqn:Ic1} have a magnitude $\mathcal{O}(1)$ and $\eta=k\cdot u/m$ is the energy parameter. (We note that $\nu$ is classical: even though we have written it in terms of the fine-structure parameter $\alpha \propto 1/\hbar$, we recall that $\eta \propto \hbar$.) In the limit of $\nu \to 0$ we see the four-velocity in \eqnref{eqn:ub} tends to the familiar expression for an electron in a plane wave \cite{Fedotov:2022ely}. The trajectory is then acquired from the velocity using $u = dx/d\tau = \eta dx/d\phi$. The nonlinear phase in \eqnref{eqn:kmu1} that is integrated over at the amplitude level, can be written as:
\bea 
k \cdot x &=& \frac{1}{2\eta s}\left[\left(1+\mbf{r}^{2}\right)\vphi+\int^{\vphi} F_{a}(\phi)\,d\phi\right] \nn \\
F_{a}(\phi) &=& 2\mbf{r}\cdot(\mbf{a}+\nu\mbf{Y})+\left(\mbf{a}+\nu\mbf{Y}\right)^{2}+ \mathcal{R}^{2}-1, \label{eqn:exp1}
\eea
where $\mbf{a}$ and $\mbf{Y}$ have only non-zero transverse components and $\mbf{r} = (s^{-1}\mbf{k}^{\perp}-\mbf{p}^{\perp})/m$ with $\mbf{p}=m\mbf{u}$, is a convenient combination of perpendicular photon and electron momenta, which for a head-on collision has a magnitude that is approximately the radiation emission angle, and $s = \vkap \cdot k / \vkap \cdot p$ is the lightfront fraction momentum of the radiation.

Without regularisation, the integral in \eqnref{eqn:kmu1} is divergent, on account of pure phase terms occurring in the integral for $\tilde{\j}(k)$. These can be removed in the standard way \cite{boca09,Ilderton:2020rgk} by introducing a regulator to perform the integration over all $\phi$ and then letting the regulator tend to zero in the finite result. We arrive at:
\bea 
\int_{-\infty}^{\infty} b(\phi)\,\mbox{e}^{i k\cdot x(\phi)}\,d\phi =
i\int_{0}^{\Phi}\,\mbox{e}^{ik\cdot x(\phi)}\,\frac{d}{d\phi}\frac{b(\phi)}{\frac{d}{d\phi} \left[k\cdot x(\phi)\right]}\,d\phi, \label{eqn:reg1}
\eea
where $b(\phi)$ is in general non-zero for all $\phi$. The outer derivative in the integrand \eqnref{eqn:reg1} cannot be converted by integration by parts into a standard regularisation factor for a finite plane wave pulse, because $F_{a}(\Phi) \neq F_{a}(0)$, i.e. there is an overall field-dependent phase shift in the electron current due to radiation reaction effects.

Finally, the differential average momentum radiated can be expressed in terms of three phase integrals:
\bea 
\frac{d^{3}\ell^{\mu}}{d^{2}k^{\perp}dk^{-}} =  \frac{1}{(2\pi)^{3}}\frac{e^{2}}{2\eta^{2}m^{2}} \frac{k^{\mu}}{k^{-}}\left\{|I_{1}|^{2}+|\mbf{I}_{2}|^{2}+\tsf{Re}\left[I_{1}I_{3}^{\ast}\right]\right\}, \label{eqn:kmu1}
\eea
\bea 
I_{1} &=& i\int^{\Phi}_{0}\mbox{e}^{i\,k\cdot x}\frac{d}{d\phi}\frac{1}{\frac{d}{d\phi} \left[k\cdot x\right]}\,d\phi \nn \\
\mbf{I}_{2} &=& \int^{\Phi}_{0}\mbox{e}^{i\,k\cdot x}\left\{\mbf{a}+\frac{\nu \pmb{\Psi'}}{\xi\Phi}+i\frac{d}{d\phi}\frac{\nu (\mbf{Y}-\pmb{\Psi'}/\xi\Phi)}{\frac{d}{d\phi} \left[k\cdot x\right]}\right\}\,d\phi\nn \\
I_{3} &=& \int^{\Phi}_{0}\mbox{e}^{i\,k\cdot x}\left[|\mbf{a}|^{2}+2\nu\,\mbf{a}\cdot\mbf{Y}+i\frac{d}{d\phi}\frac{|\nu\mbf{Y}|^{2}+\mathcal{R}^{2}-1}{\frac{d}{d\phi} \left[k\cdot x\right]}\right]\,d\phi. \nn \\ \label{eqn:In}
\eea
In the limit of no radiation reaction $\nu \to 0$, these integrals tend to others in the literature for calculating nonlinear Thomson scattering in a finite plane wave \cite{King:2020hsk}.

\subsection{IR limit}

Of particular interest for experiment, is the lightfront momentum spectrum, $du/ds = d(\vkap \cdot \ell)/d(\vkap \cdot k)$, which is a good approximation to the momentum spectrum for highly relativistic electrons. In general, the phase integrals in \eqnref{eqn:In} must be evaluated numerically. However, the low-$s$ limit of the lightfront momentum spectrum can be written in a simplified manner. Since $s \propto \vkap \cdot k = \vkap^{0}k^{0}(1-\cos\vartheta)$ where the collision angle $\vartheta$ satisfies $\vartheta=0$ for radiation emitted parallel to the plane wave propagation direction, we see that the $s\to 0$ limit is reached by parallel radiation emission and/or low radiation energies, although we refer to it here for brevity as the infra-red (IR) limit.

For the case of no radiation reaction, the IR limit is known to be \cite{DiPiazza:2017raw,Ilderton:2018nws}:  
\bea 
\frac{d\avs}{ds} \stackrel{s\to 0}{\rightarrow} \frac{s\,\alpha \xi^{2}}{2\eta}\int_{0}^{\Phi}  |\pmb{\Psi}(\phi)|^{2} d\phi.    \label{eq:Climit}
\eea
However, radiation reaction gives a finite IR limit, which originates from the regularisation terms that describe the interaction at the beginning and end of the pulse. Therefore the non-zero IR limit is a pulse envelope effect. We find:
\bea
        \lim_{s \to 0} \frac{d \avs}{d s} =
        \left(\frac{e}{2\pi}\right)^{2} \left[ \left(\frac{2 + \Gamma}{\Gamma}\right) \ln(1 + \Gamma) - 2 \right],
    \label{eq:CRRlimit}
\eea 
where 
\bea 
    \Gamma = \lim_{\varphi\to\infty} \mathcal{R}^{2}(\vphi) - 1 = \nu \lim_{\varphi\to\infty}  X(\vphi)\left[2 + \nu X(\vphi)\right].
\eea 
We note the logarithmic form of \eqnref{eq:CRRlimit} and the dependency of the prefactor on $e^2$ rather than the combination $\alpha\eta$, is similar in form to the low-energy limit of the emitted radiation derived in \cite{dipiazza.plb.2018}. If radiation reaction effects are very small, i.e. $\Gamma\ll 1$, we find $d\avs/ds \approx (\alpha/6\pi)\Gamma^{2}$.  In the opposite limit of large radiation reaction effects $\Gamma \to \infty$, we see the IR limit is logarithmic divergent {$d\avs/ds \to (\alpha/\pi)\ln(1+\Gamma)$}.

\section{Classical LMA for RR in a CP background}
Finite plane waves are often used to model the interaction of highly relativistic charges with intense laser pulses. The plane wave typically has two timescales: i) the fast timescale of the carrier frequency; ii) the slow timescale of the pulse envelope. The local monochromatic approximation (LMA) \cite{Heinzl:2020ynb} is essentially an adiabatic approach that includes the fast timescale of the background exactly, and the slow timescale asymptotically. This is achieved by writing the scattering amplitude in a similar form to that of a monochromatic wave, but still including a dependency on the pulse envelope \cite{bamber99,cain,Hartin:2018egj}. To acquire a local rate for the process, the slow timescale is approximated by performing a local expansion of the pulse envelope and integrating in the phase difference variable , keeping just leading-order term (see e.g. \cite{Heinzl:2020ynb}).

Here we outline key steps in the derivation of the LMA in a circularly-polarised background for nonlinear Thomson scattering with Landau Lifshitz radiation reaction. To do this we consider a scaled potential of the form:
\be 
\mbf{a} = m \xi g\left(\frac{\vphi}{\Phi}\right) \left\{\cos\vphi, \sin\vphi\right\}, \label{eqn:a1}
\ee 
with $|g|\leq 1$. We choose $\Phi \gg 1$ to set the pulse envelope timescale to be much slower than the carrier frequency timescale. Extending the LMA from the case of regular plane waves, requires identifying timescales of any new effects. For example, in \cite{Blackburn:2021rqm} where the LMA was applied to chirped backgrounds, a condition of applicability was found that was expressed as a `slowly-varying chirp'. The situation is not dissimilar here: as the electron slows down due to RR effects, the local frequency it experiences effectively becomes a function of phase.

The kinematics of the process are set by the form of the exponent, \eqnref{eqn:exp1}. Central to expressing the rate of the process in quasi-monochromatic form, is the Jacobi-Anger expansion \cite{landau8} e.g.:
\bea 
\mbox{e}^{-iz\sin(\vphi-\vphi_{0})} = \sum_{n=-\infty}^{\infty}J_{n}(z)\,\mbox{e}^{iz\vphi_{0}},
\eea
where $z=z(\vphi)$ depends on the pulse envelope, $g$. Using \eqnref{eqn:exp1}, we see:
\bea 
k\cdot x &=& -z\sin(\vphi-\vphi_{0})+\frac{1}{2\eta s}\left\{(1+\mbf{r}^{2})\vphi \right. \nn \\
&& \left. + \int^{\vphi}\left[\left(\mbf{a} + \nu \mbf{Y}\right)^{2}+\mathcal{R}^{2}(\phi)-1\right]d\phi \right\},
\eea  
where we find:
\bea 
z &=& \frac{sr\xi}{\eta}\Big|g\left(\frac{\phi}{\Phi}\right)\Big|\left\{\left[1+\nu \int^{\vphi}g^{2}(\phi)d\phi\right]^{2} \right. \nn \\
&& \left. +\left[\frac{\nu}{\Phi}\left(\frac{1}{\xi^{2}}+g^{2}\left(\frac{\phi}{\Phi}\right)\right)\right]^{2}\right\}^{1/2}\label{eqn:zdef}
\eea 
with 
\bea 
\tan\vphi_{0} = \frac{\left[1+\nu \int^{\vphi}g^{2}(\phi)\right]\tan\psi-\frac{\nu}{\Phi}\left[\frac{1}{\xi^{2}}+g^{2}\left(\frac{\phi}{\Phi}\right)\right]}{\left[1+\nu \int^{\vphi}g^{2}(\phi)\right]+\frac{\nu}{\Phi}\left[\frac{1}{\xi^{2}}+g^{2}\left(\frac{\phi}{\Phi}\right)\right]\tan\psi}
\label{eqn:phi0a}
\eea
where $\mbf{r} = r \{\cos\psi, \sin\psi\}$. The expressions in \eqnrefs{eqn:zdef}{eqn:phi0a}, were arrived at by applying the `slowly-varying' approximation and dropping all terms with derivatives of the envelope $g$. To justify this, consider the $[\nu X(\phi)]^{2}$ term that occurs in $\mathcal{R}^{2}(\phi)$. We see:
\[
[\nu X(\phi)]^{2} = \left[\frac{2}{3}\alpha\eta \xi^{2} \Phi \int^{\vphi} g^{2}(x) + \frac{1}{\Phi^{2}}g'^{2}(x)~dx\right]^{2}.
\]
To justify the slowly-varying approximation, we require the cross-term, which scales as $\sim (\alpha\eta\xi^{2})^{2}$ satisfy $(\alpha\eta\xi^{2})^{2} \ll 1$. This is violated when $\alpha\eta\xi^{2} \sim \mathcal{O}(1)$, i.e. if there is significant deceleration of the electron due to RR within a \emph{single cycle}. This can only be the case in the classical regime of $\eta \ll 1$ if $\xi \gg 1$, where classical RR effects can in any case be calculated with the classical locally constant field approximation with RR \cite{Heinzl:2021mji,Piazza:2021vxi}. The requirement here of $\alpha\eta\xi^{2}\ll 1$ is essentially requiring `slowly-varying RR'.

Proceeding from \eqnrefs{eqn:zdef}{eqn:phi0a}, we see more terms can be dropped in $z$ and $\tan\psi_{0}$: the brackets premultiplied by $\nu/\Phi \sim \alpha \eta \xi^{2}$ can also be neglected when $\alpha\eta\xi^{2}\ll 1$. The only possibility that the bracket is not negligible, is if $\xi^{2}\Phi \ll 1$. From other works on the LMA, it is known that pulse envelope effects can be important when $\xi$ is sufficiently small. Therefore, we demand here also that $\xi^{2}\Phi \not\ll 1$. It then follows in this limit from \eqnref{eqn:phi0a} that the azimuthal symmetry of the CP background is recovered, i.e. $\vphi_{0} = \psi$, allowing one transverse momentum integration to be performed. The Bessel argument can then be written as:
\bea 
z(\phi) = \frac{sr\xi}{\eta}\Bigg|g\left(\frac{\phi}{\Phi}\right)\Bigg| \mathcal{R}(\phi).
\eea

The rest of the derivation follows as in the simpler case of just a plane wave background with no RR (see e.g. \cite{Heinzl:2020ynb} for details). We eventually find an expression for the rate of lightfront momentum radiated, $d\avs/d\phi$, which can be written as a sum over harmonics $d\avs/d\phi = \sum_{n=1}^{\infty}d\avs_{n}/d\phi$. The lightfront momentum radiated can be written in terms of the classical equivalent number of `photons' $N_{\gamma}$ radiated:
\bea 
\avs_{n} = \int_{0}^{s_{n}} s\, \frac{dN_{\gamma,n}}{ds}\,ds, 
\eea
where, despite the increased calculational difficulty introduced by radiation reaction, the classical equivalent number of photons radiated per harmonic can be written as:
\bea 
\frac{d N_{\gamma,n}}{d\phi} &=& -\frac{e^{2}}{4\pi\eta(\phi)} 
\int_{0}^{\infty} ds(\phi) \,\int_{0}^{\infty}
d\left[r^{2}(\phi)\right]\frac{s(\phi)}{2\eta(\phi)} \nn \\
&& \times\, \delta\left\{\frac{s(\phi)}{2\eta(\phi)}\left[r^{2}(\phi)-r_{n}^{2}(\phi)\right]\right\}\left\{ J_{n}^{2}(z)+ \right.  \nn \\
&&\left.   \frac{1}{2}\xi^{2}g^{2}\left(\frac{\phi}{\Phi}\right)\left[2J_{n}^{2}\left(z\right)-J_{n+1}^{2}\left(z\right)-J_{n-1}^{2}\left(z\right)\right]\right\}, \nn \\ \label{eqn:diffRate1}
\eea
which is an identical form to the LMA rate for nonlinear Thomson scattering without RR, but now with new parameters:
\bea 
\eta(\phi)=\frac{\eta}{\mathcal{R}(\phi)}; \qquad s(\phi) = \frac{\vkap \cdot k}{\vkap\cdot p(\phi)} = s\mathcal{R}(\phi)
\eea
and 
\bea 
\mbf{r}(\phi) = \frac{\mbf{k^{\perp}}}{s(\phi)} -\mbf{p^{\perp}}(\phi) = \frac{\mbf{r}}{\mathcal{R}(\phi)},
\eea
which have been made `local' due to RR effects, giving the Bessel function arguments:
\bea 
z = \frac{s(\phi)}{\eta(\phi)} r(\phi) \xi \Big|g\left(\frac{\phi}{\Phi}\right)\Big|
\eea
with lightfront and transverse momentum harmonic values:
\bea 
s_{n}(\phi) &=& \frac{2n\eta(\phi)}{1+\xi^{2}g^{2}\left(\frac{\phi}{\Phi}\right)}; \nn \\
r_{n}^{2}(\phi) &=& \frac{2n\eta(\phi)}{s(\phi)} - \left\{1+\left[\mbf{u}^{\perp}(\phi)-\frac{\mbf{u}_{0}^{\perp}}{\mathcal{R}(\phi)}\right]^{2}\right\}; \nn \\ 
\left(\mbf{u}^{\perp}(\phi) -\frac{\mbf{u}_{0}^{\perp}}{\mathcal{R}(\phi)}\right)^{2}&=& \frac{\xi^{2}g^{2}\left(\frac{\phi}{\Phi}\right)+|\nu\mbf{Y}|^{2}}{\mathcal{R}^{2}(\phi)} \label{eqn:diffharms}
\eea
(derivatives of the pulse envelope have been neglected to acquire the final line of the above equation)
where $\mbf{a}=\mbf{a}(\phi)$ and $\mbf{Y}=\mbf{Y}(\phi)$. In the limit of $\nu \to 0$ i.e. $\eta(\phi)\to \eta$, $s(\phi)\to s$, $\mathcal{R}\to 1$, we see $r_{n}^{2}$, which defines the transverse momentum harmonic position, tends to the value without RR when one notes that the squared potential term $\mbf{a}^{2}/m$ can be phrased in terms of the change in transverse velocity, $\mbf{u^{\perp}}-\mbf{u}^{\perp}_{0}$.

The transverse momentum integral in \eqnref{eqn:diffRate1} can be straightforwardly integrated to give the lightfront spectrum:
\bea 
\frac{d N_{\gamma,n}}{d\phi} &=& -\frac{e^{2}}{4\pi\eta(\phi)} \int_{0}^{s_{n}(\phi)} ds(\phi) \left\{ J_{n}^{2}(z)+ \right.  \nn \\
&&\left.   \frac{1}{2}\xi^{2}g^{2}\left(\frac{\phi}{\Phi}\right)\left[2J_{n}^{2}\left(z\right)-J_{n+1}^{2}\left(z\right)-J_{n-1}^{2}\left(z\right)\right]\right\}, \nn \\ \label{eqn:diffRate2}
\eea
with Bessel function argument:
\bea 
z=\frac{2n\xi\left|g\left(\frac{\phi}{\Phi}\right)\right|}{\sqrt{\left[1+\xi^{2}g^{2}\left(\frac{\phi}{\Phi}\right)\right]}} \left[\frac{s(\phi)}{s_{n}(\phi)}\left(1-\frac{s(\phi)}{s_{n}(\phi)}\right)\right]^{1/2}. \label{eqn:z2}
\eea

\section{Radiation Reaction and Harmonics}
From the LMA, it is clear that the position of harmonics are influenced by radiation reaction. The end of the first harmonic range, the `Thomson edge' (the classical limit of the `Compton edge' \cite{harvey09}) is a prominent structure in particle spectra when $\xi \sim \mathcal{O}(1)$. For nonlinear Thomson (NLT), nonlinear Thomson with RR and nonlinear Compton, we have the `instanteous' first harmonic edges:
\bea 
s_{1}^{\tsf{\tiny NLT}}(\phi) = \frac{2\eta}{1+\xi^{2}(\phi)}; \quad s_{1}^{\tsf{\tiny NLT $+$ RR}}(\phi) = \frac{s_{1}^{\tsf{\tiny NLT}}(\phi)}{\mathcal{R}^{2}(\phi)} \label{eqn:s1s}
\eea
and $s_{1}^{\tsf{\tiny QED}}(\phi) = s_{1}^{\tsf{\tiny NLT}}(\phi)/[1+s_{1}^{\tsf{\tiny NLT}}(\phi)]$. The Thomson edge in the integrated spectrum is then given by the minimum of these values, e.g. $s_{1}^{\tsf{\tiny NLT}} = \min_{\phi}s_{1}^{\tsf{\tiny NLT}}(\phi) = 2\eta/(1+\xi^{2})$. We see that already when $\mathcal{R}^{2}(\phi)-1 \sim \mathcal{O}(s_{1}^{\tsf{\tiny NLT}})$, or equivalently $\nu \sim \mathcal{O}[\eta/(1+\xi^{2})] \ll 1$, classical RR effects are as large as QED effects and should be included in analysis of experiments seeking to measure harmonic structure. Therefore in terms of the spectrum, it is not necessary that $\nu \sim \mathcal{O}(1)$ for RR effects to become evident: it suffices that $\nu$ is of the order of the energy resolution of the measurement, e.g. $\nu \sim \mathcal{O}(0.1) \ll 1$. 

\begin{figure}[h!!]
\centering
\includegraphics[width=0.99\linewidth]{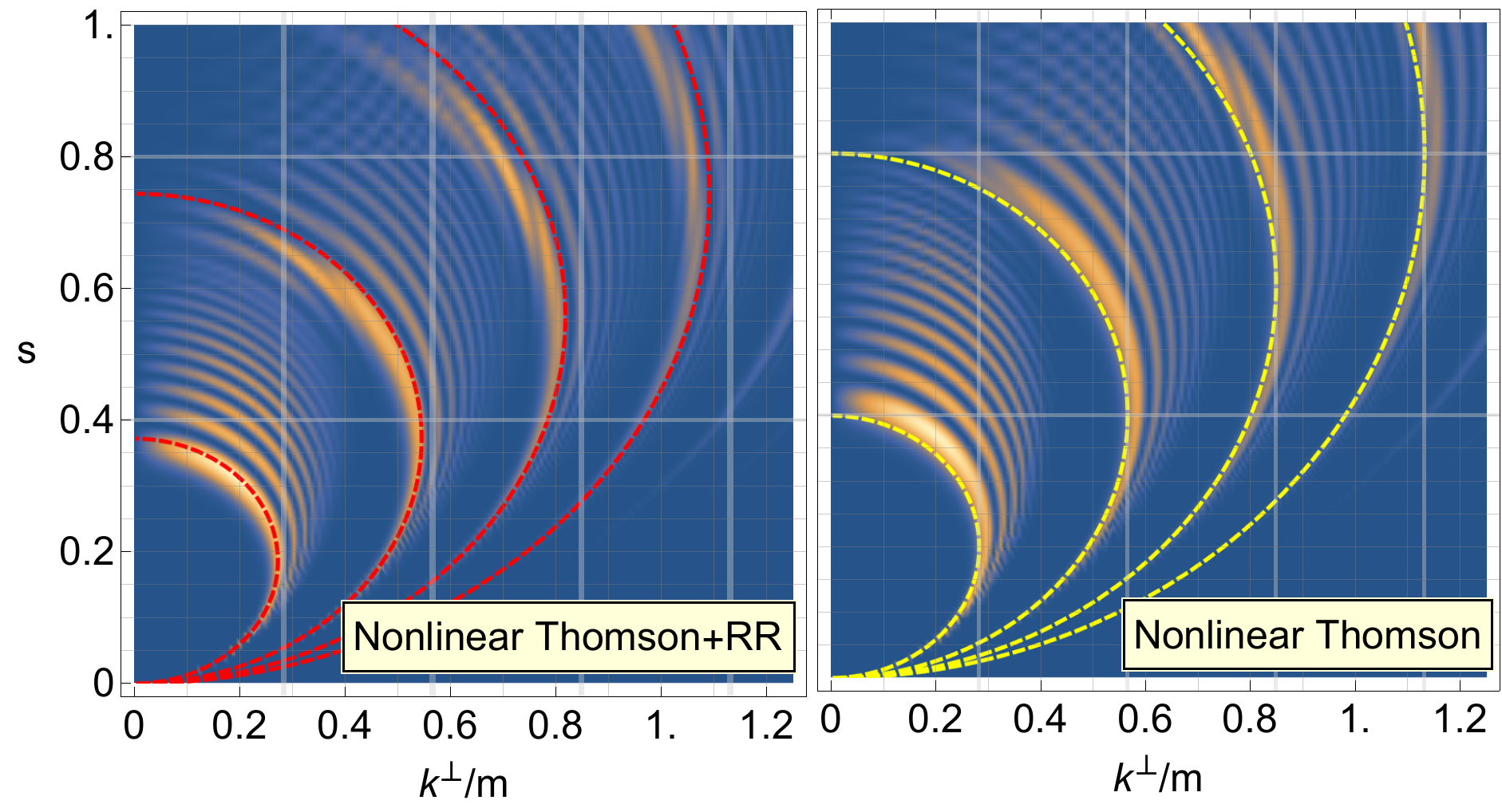}
\caption{Plots of $d^{2}\avs/ds \, d(k^{\perp}/m)$ for $\xi=1$, $\eta=0.4$, $N=32$ with (left) and without (right) RR. The dashed lines indicate the approximate position of the harmonic edge in variables $(k^{\perp}/m,s)$ with RR (left) and without RR (right). In both plots, the bolder gridlines denote the harmonic edges  without RR in the the integrated transverse momentum ($k^{\perp}/m$, vertical) and lightfront fraction ($s$, horizontal) variables. } \label{fig:plots1}
\end{figure}

To demonstrate this point and benchmark the LMA with numerical evaluation of the exact plane-wave result, in \figref{fig:plots1} we present results for the double-differential spectrum from nonlinear Thomson scattering with and without RR for $\xi=1$, $\eta=0.4$ and $N=32$ for a sine-squared pulse envelope, i.e. choosing in \eqnref{eqn:a1} that $g(\vphi/\Phi) = \sin^{2}(\pi\vphi/\Phi)$ for $0<\vphi<\Phi$ and $g=0$.
It can be seen from \figref{fig:plots1} how RR causes a red-shifting of the lower harmonic edge and a greater overlap of the kinematic ranges for low harmonics even at $\xi=1$, which is usually only a feature of $\xi>1$ spectra \cite{Heinzl:2009nd}. The harmonic kinematic range indicated in \figref{fig:plots1} is simply where the delta function in \eqnref{eqn:diffRate1} has support given in \eqnref{eqn:diffharms}. The case $\xi \gg 1$, where harmonics are no longer visible but classical RR changes the kinematic range, has also be analysed in the literature \cite{DiPiazza:2009zz,thomas.prx.2012,Heinzl:2013txd}.

\begin{figure}[h!!]
\centering
\includegraphics[width=0.8\linewidth]{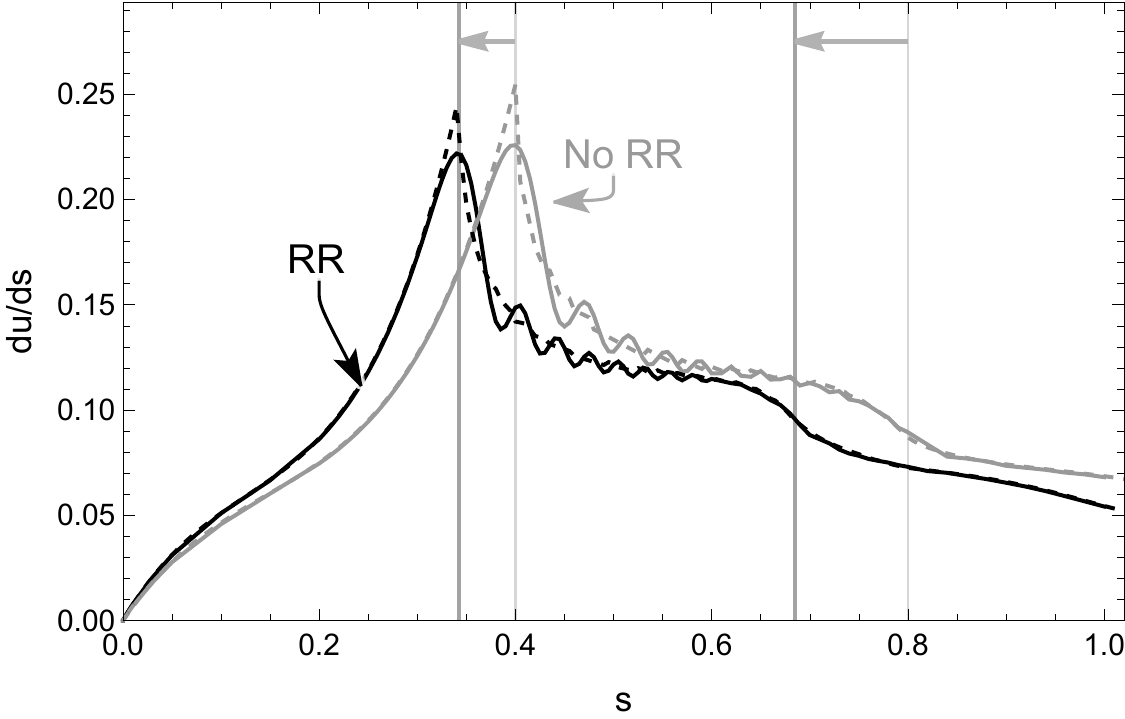}\\
\includegraphics[width=0.8\linewidth]{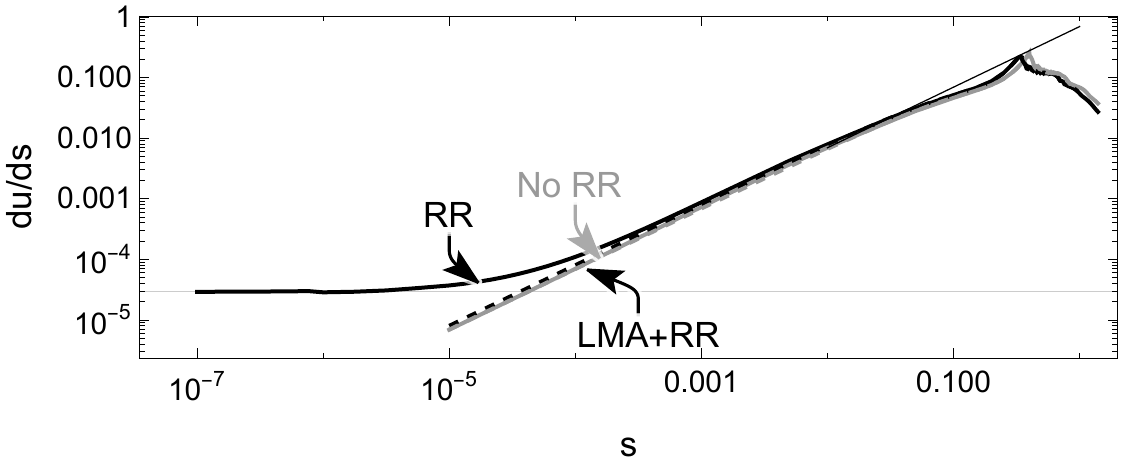}
\caption{Lightfront momentum spectra for $\xi=1$, $\eta=0.4$, $N=32$, corresponding to a RR parameter $\nu=0.39$. Top plot: numerical evaluation of lightfront momentum spectrum of emitted radiation for: exact plane wave with RR (dark solid line); without RR (light solid line); LMA with RR (dark dashed line); LMA without RR (light dashed line). The harmonic positions are given by calculating the minimum with respect to $\phi$ of the `instantaneous' harmonic positions in \eqnref{eqn:s1s}. Bottom plot: The low-$s$ limit is non-zero for the exact plane wave with RR result, which is missed by the LMA with RR. The gridline correponds to evaluation of equation \eqnref{eq:CRRlimit}.} \label{fig:plots2}
\end{figure}

To make the Thomson edge clearer, we integrate the data in \figref{fig:plots1} over the transverse momentum variable $r$, to give the lightfront momentum spectrum, $d\avs/ds$, plotted in \figref{fig:plots2}. We also plot the application of the LMA formula and find excellent agreement with the exact plane wave result. The LMA as expected, correctly models the red-shifting of the harmonics due to RR, and averages through the sub-harmonic structure. One limitation of the LMA is that the low-$s$ limit of the spectrum tends to zero, whereas the exact result tends to a finite non-zero value. In this parameter region, RR effects involving the pulse envelope dominate. It has been already established \cite{King:2020hsk,Tang:2021qht} that the LMA cannot capture interference effects on the length scale of the pulse envelope, due to it only including local variations in the carrier frequency amplitude. To ascertain when the non-zero low-$s$ limit due to RR is important, one can use the value of $s=s_{\ast}$ at which the non-RR and RR low-$s$ limits in \eqnrefs{eq:Climit}{eq:CRRlimit} intersect. This leads to: $k^{0}/m > \eta s_{\ast}/2(\vkap^{0}/m)$, and for the results in \figref{fig:plots2}, for an optical laser frequency, the IR effects become important when the  radiation has a frequency on the order of an MeV.

\begin{figure}[h!!]
\centering
\includegraphics[width=0.8\linewidth]{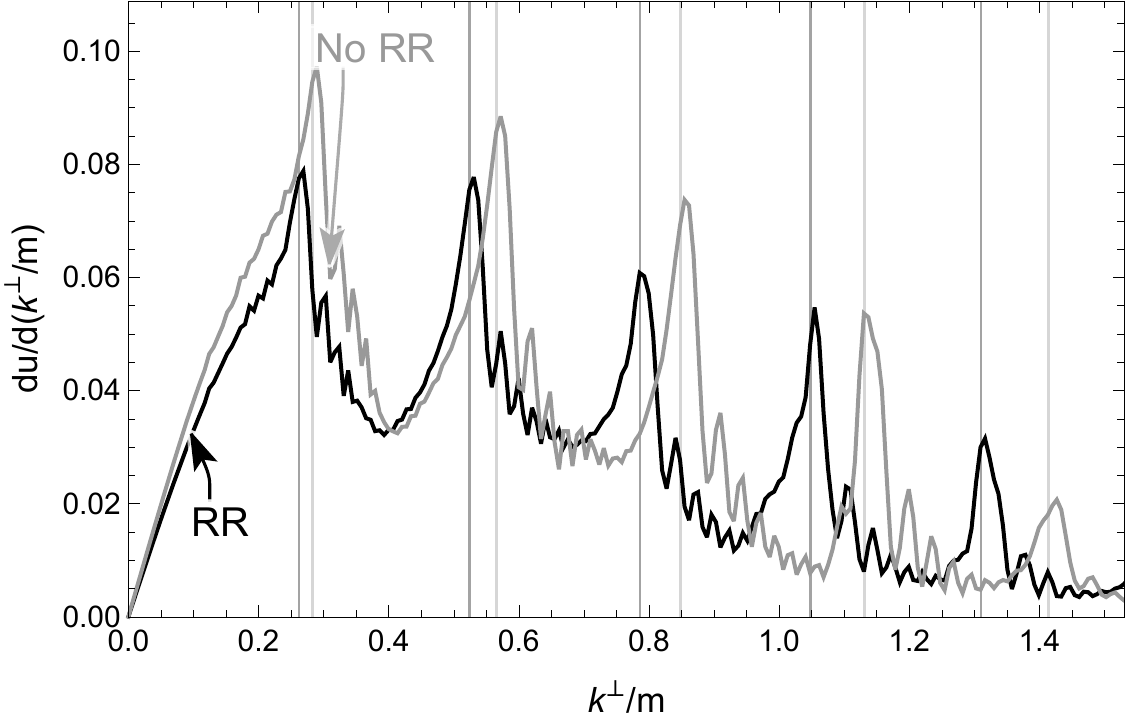}
\caption{The differential spectrum $d\avs/d(k^{\perp}/m)$. The darker (lighter) gridlines correspond to the predicted harmonic position from the LMA with (without) RR. The harmonic positions are given by calculating the minimum with respect to $\phi$ of the `instantaneous' harmonic positions in \eqnref{eqn:rns}} \label{fig:plots3}
\end{figure}

We can also observe red-shifting of harmonics in the transverse momentum spectrum by integrating the data in \figref{fig:plots1} over the lightfront momentum variable, $s$, to give the spectrum $d\avs/d(k^{\perp}/m)$ shown in \figref{fig:plots3}.
Good agreement is shown with the predicted position of the transverse momentum harmonics:
\bea 
k^{\perp}_{n}=\trm{min}_{\phi}k^{\perp}_{n}(\phi); \quad k^{\perp}_{n}(\phi)=\frac{nm\eta(\phi)}{\sqrt{1+\xi^{2}(\phi)}}, \label{eqn:rns}
\eea
which we see are redshifted similar to in the lightfront momentum spectrum with higher harmonics being shifted by a greater amount than lower harmonics.

\section{Summary}
The radiation spectrum has been calculated for an intense finite plane wave colliding with an electron subject to the Landau-Lifshitz radiation reaction force. Numerical evaluation of the exact result has been compared with a locally monochromatic approximation (LMA) and good agreement found in the red-shifting of harmonics compared to the case of no radation reaction (nonlinear Thomson scattering). 

Radiation reaction has been \cite{Wistisen:2017pgr,cole18,Poder:2017dpw} and continues to be searched for in laser-particle experiments. Measuring harmonic structure is useful in experiments to assess the nonlinearity of interaction \cite{Chen1998a,khrennikov15,sakai15} and is a target of future experiments \cite{Fleck:2020opg,Abramowicz:2021zja}. Analysing the LMA here gave simple formulas for predicting the position of harmonic edges in outgoing particle spectra and we suggest that the position of these edges can be used as a probe of radiation reaction. The LMA presented in this letter has now been implemented in the open source Monte Carlo numerical simulation code Ptarmigan \cite{ptarmiganPaper,ptarmigan} to aid in this search.  
 
This classical approach, which includes, in some limit, an all-order interaction of the charge with its own radiation field \cite{Torgrimsson:2021wcj}, is most useful for experiments employing long pulses (so that charges radiate many times) where the intensity parameter is not much greater than unity (so that the harmonic structure is clear) and where the energy and strong-field parameters are much smaller than unity (so that quantum effects are small).

In the spectrum of emitted particles, classical RR becomes important when the parameter $\nu=2\alpha\eta\xi^{2}\Phi/3$ is of the order of the resolution of the energy measurement, e.g. already at $\nu\sim \mathcal{O}(0.1)$, classical RR effects should be apparent.

\section*{Acknowledgments}
The author thanks A. Ilderton and T. Blackburn for useful discussions and reading the manuscript.

\bibliography{masterRR}

\providecommand{\noopsort}[1]{}
\begin{thebibliography}{10}
\expandafter\ifx\csname url\endcsname\relax
  \def\url#1{\texttt{#1}}\fi
\expandafter\ifx\csname urlprefix\endcsname\relax\def\urlprefix{URL }\fi
\expandafter\ifx\csname href\endcsname\relax
  \def\href#1#2{#2} \def\path#1{#1}\fi

\bibitem{Burton:2014wsa}
D.~A. Burton, A.~Noble, {Aspects of electromagnetic radiation reaction in
  strong fields}, Contemp. Phys. 55~(2) (2014) 110--121.
\newblock \href {http://arxiv.org/abs/1409.7707} {\path{arXiv:1409.7707}},
  \href {https://doi.org/10.1080/00107514.2014.886840}
  {\path{doi:10.1080/00107514.2014.886840}}.

\bibitem{Blackburn:2019rfv}
T.~G. Blackburn, {Radiation reaction in electron-beam interactions with
  high-intensity lasers}, Plasma Phys. 4 (2020) 5.
\newblock \href {http://arxiv.org/abs/1910.13377} {\path{arXiv:1910.13377}},
  \href {https://doi.org/10.1007/s41614-020-0042-0}
  {\path{doi:10.1007/s41614-020-0042-0}}.

\bibitem{Gonoskov:2021hwf}
A.~Gonoskov, T.~G. Blackburn, M.~Marklund, S.~S. Bulanov, {Charged particle
  motion and radiation in strong electromagnetic fields}, Rev. Mod. Phys.
  94~(4) (2022) 045001.
\newblock \href {http://arxiv.org/abs/2107.02161} {\path{arXiv:2107.02161}},
  \href {https://doi.org/10.1103/RevModPhys.94.045001}
  {\path{doi:10.1103/RevModPhys.94.045001}}.

\bibitem{Krivitsky:1991vt}
V.~S. Krivitsky, V.~N. Tsytovich, {Average radiation reaction force in quantum
  electrodynamics}, Sov. Phys. Usp. 34 (1991) 250--258.
\newblock \href {https://doi.org/10.1070/PU1991v034n03ABEH002352}
  {\path{doi:10.1070/PU1991v034n03ABEH002352}}.

\bibitem{Higuchi:2002qc}
A.~Higuchi, {Radiation reaction in quantum field theory}, Phys. Rev. D 66
  (2002) 105004, [Erratum: Phys.Rev.D 69, 129903 (2004)].
\newblock \href {http://arxiv.org/abs/quant-ph/0208017}
  {\path{arXiv:quant-ph/0208017}}, \href
  {https://doi.org/10.1103/PhysRevD.66.105004}
  {\path{doi:10.1103/PhysRevD.66.105004}}.

\bibitem{Ilderton:2013dba}
A.~Ilderton, G.~Torgrimsson, {Radiation reaction from QED: lightfront
  perturbation theory in a plane wave background}, Phys. Rev. D 88~(2) (2013)
  025021.
\newblock \href {http://arxiv.org/abs/1304.6842} {\path{arXiv:1304.6842}},
  \href {https://doi.org/10.1103/PhysRevD.88.025021}
  {\path{doi:10.1103/PhysRevD.88.025021}}.

\bibitem{Ilderton:2013tb}
A.~Ilderton, G.~Torgrimsson, {Radiation reaction in strong field QED}, Phys.
  Lett. B 725 (2013) 481.
\newblock \href {http://arxiv.org/abs/1301.6499} {\path{arXiv:1301.6499}},
  \href {https://doi.org/10.1016/j.physletb.2013.07.045}
  {\path{doi:10.1016/j.physletb.2013.07.045}}.

\bibitem{Torgrimsson:2020gws}
G.~Torgrimsson, {Loops and polarization in strong-field QED}, New J. Phys.
  23~(6) (2021) 065001.
\newblock \href {http://arxiv.org/abs/2012.12701} {\path{arXiv:2012.12701}},
  \href {https://doi.org/10.1088/1367-2630/abf274}
  {\path{doi:10.1088/1367-2630/abf274}}.

\bibitem{Torgrimsson:2021wcj}
G.~Torgrimsson, {Resummation of Quantum Radiation Reaction in Plane Waves},
  Phys. Rev. Lett. 127~(11) (2021) 111602.
\newblock \href {http://arxiv.org/abs/2102.11346} {\path{arXiv:2102.11346}},
  \href {https://doi.org/10.1103/PhysRevLett.127.111602}
  {\path{doi:10.1103/PhysRevLett.127.111602}}.

\bibitem{Ekman:2021vwg}
R.~Ekman, T.~Heinzl, A.~Ilderton, {Exact solutions in radiation reaction and
  the radiation-free direction}, New J. Phys. 23~(5) (2021) 055001.
\newblock \href {http://arxiv.org/abs/2102.11843} {\path{arXiv:2102.11843}},
  \href {https://doi.org/10.1088/1367-2630/abfab2}
  {\path{doi:10.1088/1367-2630/abfab2}}.

\bibitem{Ekman:2021eqc}
R.~Ekman, T.~Heinzl, A.~Ilderton, {Reduction of order, resummation, and
  radiation reaction}, Phys. Rev. D 104~(3) (2021) 036002.
\newblock \href {http://arxiv.org/abs/2105.01640} {\path{arXiv:2105.01640}},
  \href {https://doi.org/10.1103/PhysRevD.104.036002}
  {\path{doi:10.1103/PhysRevD.104.036002}}.

\bibitem{dipiazza08LL}
A.~{Di Piazza}, \href{https://doi.org/10.1007/s11005-008-0228-9}{Exact solution
  of the landau-lifshitz equation in a plane wave'}, Lett Math Phys 83 (2008)
  305–313.
\newblock \href {https://doi.org/10.1007/s11005-008-0228-9}
  {\path{doi:10.1007/s11005-008-0228-9}}.
\newline\urlprefix\url{https://doi.org/10.1007/s11005-008-0228-9}

\bibitem{dipiazza12}
A.~{Di Piazza}, et~al., Extremely high-intensity laser interactions with
  fundamental quantum systems, Rev. Mod. Phys. 84 (2012) 1177--1228.

\bibitem{narozhny15}
N.~B. Narozhny, A.~M. Fedotov, Extreme light physics, Contemporary Physics 56
  (2015) 249--268.

\bibitem{Fedotov:2022ely}
A.~Fedotov, A.~Ilderton, F.~Karbstein, B.~King, D.~Seipt, H.~Taya,
  G.~Torgrimsson, {Advances in QED with intense background fields}, Phys. Rept.
  1010 (2023) 1--138.
\newblock \href {http://arxiv.org/abs/2203.00019} {\path{arXiv:2203.00019}},
  \href {https://doi.org/10.1016/j.physrep.2023.01.003}
  {\path{doi:10.1016/j.physrep.2023.01.003}}.

\bibitem{Seipt:2020diz}
D.~Seipt, B.~King, {Spin- and polarization-dependent
  locally-constant-field-approximation rates for nonlinear Compton and
  Breit-Wheeler processes}, Phys. Rev. A 102~(5) (2020) 052805.
\newblock \href {http://arxiv.org/abs/2007.11837} {\path{arXiv:2007.11837}},
  \href {https://doi.org/10.1103/PhysRevA.102.052805}
  {\path{doi:10.1103/PhysRevA.102.052805}}.

\bibitem{Neitz:2013qba}
N.~Neitz, A.~Di~Piazza, {Stochasticity Effects in Quantum Radiation Reaction},
  Phys. Rev. Lett. 111~(5) (2013) 054802.
\newblock \href {http://arxiv.org/abs/1301.5524} {\path{arXiv:1301.5524}},
  \href {https://doi.org/10.1103/PhysRevLett.111.054802}
  {\path{doi:10.1103/PhysRevLett.111.054802}}.

\bibitem{yoffe.njp.2015}
S.~R. Yoffe, Y.~Kravets, A.~Noble, D.~A. Jaroszynski,
  \href{https://iopscience.iop.org/article/10.1088/1367-2630/17/5/053025
  https://iopscience.iop.org/article/10.1088/1367-2630/17/5/053025/meta}{{Longitudinal
  and transverse cooling of relativistic electron beams in intense laser
  pulses}}, New Journal of Physics 17~(5) (2015) 053025.
\newblock \href {https://doi.org/10.1088/1367-2630/17/5/053025}
  {\path{doi:10.1088/1367-2630/17/5/053025}}.
\newline\urlprefix\url{https://iopscience.iop.org/article/10.1088/1367-2630/17/5/053025
  https://iopscience.iop.org/article/10.1088/1367-2630/17/5/053025/meta}

\bibitem{shen.prl.1972}
C.~S. Shen, D.~White,
  \href{https://link.aps.org/doi/10.1103/PhysRevLett.28.455}{{Energy Straggling
  and Radiation Reaction for Magnetic Bremsstrahlung}}, Physical Review Letters
  28~(7) (1972) 455--459.
\newblock \href {https://doi.org/10.1103/PhysRevLett.28.455}
  {\path{doi:10.1103/PhysRevLett.28.455}}.
\newline\urlprefix\url{https://link.aps.org/doi/10.1103/PhysRevLett.28.455}

\bibitem{Blackburn:2014cig}
T.~G. Blackburn, C.~P. Ridgers, J.~G. Kirk, A.~R. Bell, {Quantum radiation
  reaction in laser-electron beam collisions}, Phys. Rev. Lett. 112 (2014)
  015001.
\newblock \href {http://arxiv.org/abs/1503.01009} {\path{arXiv:1503.01009}},
  \href {https://doi.org/10.1103/PhysRevLett.112.015001}
  {\path{doi:10.1103/PhysRevLett.112.015001}}.

\bibitem{Harvey:2016uiy}
C.~Harvey, A.~Gonoskov, A.~Ilderton, M.~Marklund, {Quantum quenching of
  radiation losses in short laser pulses}, Phys. Rev. Lett. 118~(10) (2017)
  105004.
\newblock \href {http://arxiv.org/abs/1606.08250} {\path{arXiv:1606.08250}},
  \href {https://doi.org/10.1103/PhysRevLett.118.105004}
  {\path{doi:10.1103/PhysRevLett.118.105004}}.

\bibitem{Wistisen:2017pgr}
T.~N. Wistisen, A.~Di~Piazza, H.~V. Knudsen, U.~I. Uggerh\o{}j, {Experimental
  evidence of quantum radiation reaction in aligned crystals}, Nature Commun.
  9~(1) (2018) 795.
\newblock \href {http://arxiv.org/abs/1704.01080} {\path{arXiv:1704.01080}},
  \href {https://doi.org/10.1038/s41467-018-03165-4}
  {\path{doi:10.1038/s41467-018-03165-4}}.

\bibitem{cole18}
J.~M. Cole, K.~T. Behm, E.~Gerstmayr, T.~G. Blackburn, J.~C. Wood, C.~D. Baird,
  M.~J. Duff, C.~Harvey, A.~Ilderton, A.~S. Joglekar, K.~Krushelnick,
  S.~Kuschel, M.~Marklund, P.~McKenna, C.~D. Murphy, K.~Poder, C.~P. Ridgers,
  G.~M. Samarin, G.~Sarri, D.~R. Symes, A.~G.~R. Thomas, J.~Warwick, M.~Zepf,
  Z.~Najmudin, S.~P.~D. Mangles,
  \href{https://link.aps.org/doi/10.1103/PhysRevX.8.011020}{Experimental
  evidence of radiation reaction in the collision of a high-intensity laser
  pulse with a laser-wakefield accelerated electron beam}, Phys. Rev. X 8
  (2018) 011020.
\newblock \href {https://doi.org/10.1103/PhysRevX.8.011020}
  {\path{doi:10.1103/PhysRevX.8.011020}}.
\newline\urlprefix\url{https://link.aps.org/doi/10.1103/PhysRevX.8.011020}

\bibitem{Poder:2017dpw}
K.~Poder, et~al., {Experimental Signatures of the Quantum Nature of Radiation
  Reaction in the Field of an Ultraintense Laser}, Phys. Rev. X 8~(3) (2018)
  031004.
\newblock \href {http://arxiv.org/abs/1709.01861} {\path{arXiv:1709.01861}},
  \href {https://doi.org/10.1103/PhysRevX.8.031004}
  {\path{doi:10.1103/PhysRevX.8.031004}}.

\bibitem{Abramowicz:2021zja}
H.~Abramowicz, et~al., {Conceptual design report for the LUXE experiment}, Eur.
  Phys. J. ST 230~(11) (2021) 2445--2560.
\newblock \href {http://arxiv.org/abs/2102.02032} {\path{arXiv:2102.02032}},
  \href {https://doi.org/10.1140/epjs/s11734-021-00249-z}
  {\path{doi:10.1140/epjs/s11734-021-00249-z}}.

\bibitem{samarin18}
G.~M. Samarin, M.~Zepf, G.~Sarri,
  \href{https://doi.org/10.1080/09500340.2017.1353655}{Radiation reaction
  studies in an all-optical set-up: experimental limitations}, Journal of
  Modern Optics 65~(11) (2018) 1362--1369.
\newblock \href
  {http://arxiv.org/abs/https://doi.org/10.1080/09500340.2017.1353655}
  {\path{arXiv:https://doi.org/10.1080/09500340.2017.1353655}}, \href
  {https://doi.org/10.1080/09500340.2017.1353655}
  {\path{doi:10.1080/09500340.2017.1353655}}.
\newline\urlprefix\url{https://doi.org/10.1080/09500340.2017.1353655}

\bibitem{danson19}
C.~N. Danson, C.~Haefner, J.~Bromage, T.~Butcher, J.-C.~F. Chanteloup, E.~A.
  Chowdhury, A.~Galvanauskas, L.~A. Gizzi, J.~Hein, D.~I. Hillier, et~al.,
  Petawatt and exawatt class lasers worldwide, High Power Laser Science and
  Engineering 7 (2019) e54.
\newblock \href {https://doi.org/10.1017/hpl.2019.36}
  {\path{doi:10.1017/hpl.2019.36}}.

\bibitem{Fedotov:2017conjecture}
A.~M. Fedotov, {Conjecture of perturbative QED breakdown at $\alpha\chi^{2/3}
  \gtrsim 1$}, J. Phys. Conf. Ser. 826~(1) (2017) 012027.
\newblock \href {http://arxiv.org/abs/1608.02261} {\path{arXiv:1608.02261}},
  \href {https://doi.org/10.1088/1742-6596/826/1/012027}
  {\path{doi:10.1088/1742-6596/826/1/012027}}.

\bibitem{Podszus:2018hnz}
T.~Podszus, A.~Di~Piazza, {High-energy behavior of strong-field QED in an
  intense plane wave}, Phys. Rev. D 99~(7) (2019) 076004.
\newblock \href {http://arxiv.org/abs/1812.08673} {\path{arXiv:1812.08673}},
  \href {https://doi.org/10.1103/PhysRevD.99.076004}
  {\path{doi:10.1103/PhysRevD.99.076004}}.

\bibitem{Ilderton:2019kqp}
A.~Ilderton, {Note on the conjectured breakdown of QED perturbation theory in
  strong fields}, Phys. Rev. D 99~(8) (2019) 085002.
\newblock \href {http://arxiv.org/abs/1901.00317} {\path{arXiv:1901.00317}},
  \href {https://doi.org/10.1103/PhysRevD.99.085002}
  {\path{doi:10.1103/PhysRevD.99.085002}}.

\bibitem{Heinzl:2021mji}
T.~Heinzl, A.~Ilderton, B.~King, {Classical Resummation and Breakdown of
  Strong-Field QED}, Phys. Rev. Lett. 127~(6) (2021) 061601.
\newblock \href {http://arxiv.org/abs/2101.12111} {\path{arXiv:2101.12111}},
  \href {https://doi.org/10.1103/PhysRevLett.127.061601}
  {\path{doi:10.1103/PhysRevLett.127.061601}}.

\bibitem{Piazza:2021vxi}
A.~D. Piazza, G.~Audagnotto, {Analytical spectrum of nonlinear Thomson
  scattering including radiation reaction}, Phys. Rev. D 104~(1) (2021) 016007.
\newblock \href {http://arxiv.org/abs/2102.11260} {\path{arXiv:2102.11260}},
  \href {https://doi.org/10.1103/PhysRevD.104.016007}
  {\path{doi:10.1103/PhysRevD.104.016007}}.

\bibitem{boca09}
M.~Boca, V.~Florescu,
  \href{https://link.aps.org/doi/10.1103/PhysRevA.80.053403}{Nonlinear compton
  scattering with a laser pulse}, Phys. Rev. A 80 (2009) 053403.
\newblock \href {https://doi.org/10.1103/PhysRevA.80.053403}
  {\path{doi:10.1103/PhysRevA.80.053403}}.
\newline\urlprefix\url{https://link.aps.org/doi/10.1103/PhysRevA.80.053403}

\bibitem{dinu12}
V.~Dinu, T.~Heinzl, A.~Ilderton, Infrared divergences in plane wave
  backgrounds, Phys. Rev. D 86 (2012) 085037.

\bibitem{harvey09}
C.~Harvey, T.~Heinzl, A.~Ilderton, Signatures of high-intensity compton
  scattering, Phys. Rev. A 79 (2009) 063407.

\bibitem{Heinzl:2020ynb}
T.~Heinzl, B.~King, A.~J. Macleod, {The locally monochromatic approximation to
  QED in intense laser fields}, Phys. Rev. A 102 (2020) 063110.
\newblock \href {http://arxiv.org/abs/2004.13035} {\path{arXiv:2004.13035}},
  \href {https://doi.org/10.1103/PhysRevA.102.063110}
  {\path{doi:10.1103/PhysRevA.102.063110}}.

\bibitem{ptarmiganPaper}
T.~G. Blackburn, B.~King, S.~Tang, Simulations of laser-driven strong-field qed
  with ptarmigan: Resolving wavelength-scale interference and $\gamma$-ray
  polarization (2023).
\newblock \href {http://arxiv.org/abs/2305.13061} {\path{arXiv:2305.13061}}.

\bibitem{ptarmigan}
T.~G. Blackburn, \href{https://github.com/tgblackburn/ptarmigan}{Ptarmigan},
  Github repository (2023).
\newline\urlprefix\url{https://github.com/tgblackburn/ptarmigan}

\bibitem{hadad.prd.2010}
Y.~Hadad, L.~Labun, J.~Rafelski, N.~Elkina, C.~Klier, H.~Ruhl,
  \href{https://link.aps.org/doi/10.1103/PhysRevD.82.096012}{{Effects of
  radiation reaction in relativistic laser acceleration}}, Physical Review D
  82~(9) (2010) 96012.
\newblock \href {https://doi.org/10.1103/PhysRevD.82.096012}
  {\path{doi:10.1103/PhysRevD.82.096012}}.
\newline\urlprefix\url{https://link.aps.org/doi/10.1103/PhysRevD.82.096012}

\bibitem{schlegel.njp.2012}
T.~Schlegel, V.~T. Tikhonchuk, {Classical radiation effects on relativistic
  electrons in ultraintense laser fields with circular polarization}, New
  Journal of Physics 14 (2012) 73034.
\newblock \href {https://doi.org/10.1088/1367-2630/14/7/073034}
  {\path{doi:10.1088/1367-2630/14/7/073034}}.

\bibitem{thomas.prx.2012}
A.~G.~R. Thomas, C.~P. Ridgers, S.~S. Bulanov, B.~J. Griffin, S.~P.~D. Mangles,
  \href{https://link.aps.org/doi/10.1103/PhysRevX.2.041004}{Strong
  radiation-damping effects in a gamma-ray source generated by the interaction
  of a high-intensity laser with a wakefield-accelerated electron beam},
  Physical Review X 2 (2012) 041004.
\newblock \href {https://doi.org/10.1103/PhysRevX.2.041004}
  {\path{doi:10.1103/PhysRevX.2.041004}}.
\newline\urlprefix\url{https://link.aps.org/doi/10.1103/PhysRevX.2.041004}

\bibitem{vranic.prl.2014}
M.~Vranic, J.~L. Martins, J.~Vieira, R.~A. Fonseca, L.~O. Silva,
  \href{https://journals.aps.org/prl/abstract/10.1103/PhysRevLett.113.134801}{{All-optical
  radiation reaction at 1021 W/cm2}}, Physical Review Letters 113~(13) (2014)
  134801.
\newblock \href {https://doi.org/10.1103/PhysRevLett.113.134801}
  {\path{doi:10.1103/PhysRevLett.113.134801}}.
\newline\urlprefix\url{https://journals.aps.org/prl/abstract/10.1103/PhysRevLett.113.134801}

\bibitem{vranic.cpc.2016}
M.~Vranic, J.~L. Martins, R.~A. Fonseca, L.~O. Silva,
  \href{http://www.sciencedirect.com/science/article/pii/S001046551630090X
  http://linkinghub.elsevier.com/retrieve/pii/S001046551630090X}{Classical
  radiation reaction in particle-in-cell simulations}, Computer Physics
  Communications 204 (2016) 141--151.
\newblock \href {https://doi.org/10.1016/j.cpc.2016.04.002}
  {\path{doi:10.1016/j.cpc.2016.04.002}}.
\newline\urlprefix\url{http://www.sciencedirect.com/science/article/pii/S001046551630090X
  http://linkinghub.elsevier.com/retrieve/pii/S001046551630090X}

\bibitem{King:2020hsk}
B.~King, {Interference effects in nonlinear Compton scattering due to pulse
  envelope}, Phys. Rev. D 103~(3) (2021) 036018.
\newblock \href {http://arxiv.org/abs/2012.05920} {\path{arXiv:2012.05920}},
  \href {https://doi.org/10.1103/PhysRevD.103.036018}
  {\path{doi:10.1103/PhysRevD.103.036018}}.

\bibitem{Heintzmann:1972mn}
H.~Heintzmann, M.~Grewing, {Acceleration of charged particles and
  radiation-reaction in strong plane and spherical waves}, Z. Phys. 251 (1972)
  77--86.
\newblock \href {https://doi.org/10.1007/BF01386985}
  {\path{doi:10.1007/BF01386985}}.

\bibitem{Ilderton:2020rgk}
A.~Ilderton, A.~J. MacLeod, {The analytic structure of amplitudes on
  backgrounds from gauge invariance and the infra-red}, JHEP 04 (2020) 078.
\newblock \href {http://arxiv.org/abs/2001.10553} {\path{arXiv:2001.10553}},
  \href {https://doi.org/10.1007/JHEP04(2020)078}
  {\path{doi:10.1007/JHEP04(2020)078}}.

\bibitem{DiPiazza:2017raw}
A.~Di~Piazza, M.~Tamburini, S.~Meuren, C.~Keitel, {Implementing nonlinear
  Compton scattering beyond the local constant field approximation}, Phys. Rev.
  A 98~(1) (2018) 012134.
\newblock \href {http://arxiv.org/abs/1708.08276} {\path{arXiv:1708.08276}},
  \href {https://doi.org/10.1103/PhysRevA.98.012134}
  {\path{doi:10.1103/PhysRevA.98.012134}}.

\bibitem{Ilderton:2018nws}
A.~Ilderton, B.~King, D.~Seipt, {Extended locally constant field approximation
  for nonlinear Compton scattering}, Phys. Rev. A 99~(4) (2019) 042121.
\newblock \href {http://arxiv.org/abs/1808.10339} {\path{arXiv:1808.10339}},
  \href {https://doi.org/10.1103/PhysRevA.99.042121}
  {\path{doi:10.1103/PhysRevA.99.042121}}.

\bibitem{dipiazza.plb.2018}
A.~Di~Piazza, Analytical infrared limit of nonlinear {Thomson} scattering
  including radiation reaction, Phys. Lett. B 782 (2018) 559--565.
\newblock \href {http://arxiv.org/abs/arXiv:1804.01160}
  {\path{arXiv:arXiv:1804.01160}}, \href
  {https://doi.org/10.1016/j.physletb.2018.05.081}
  {\path{doi:10.1016/j.physletb.2018.05.081}}.

\bibitem{bamber99}
C.~Bamber, et~al., Phys. Rev. D 60 (1999) 092004.

\bibitem{cain}
P.~Chen, G.~Horton-Smith, T.~Ohgaki, A.~W. Weidemann, K.~Yokoya, {CAIN:
  Conglomérat d'ABEL et d'Interactions Non-linéaires}, Nucl. Instrum. Methods
  Phys. Res. A 355~(1) (1995) 107--110.
\newblock \href {https://doi.org/10.1016/0168-9002(94)01186-9}
  {\path{doi:10.1016/0168-9002(94)01186-9}}.

\bibitem{Hartin:2018egj}
A.~Hartin, {Strong field QED in lepton colliders and electron/laser
  interactions}, Int. J. Mod. Phys. A 33~(13) (2018) 1830011.
\newblock \href {http://arxiv.org/abs/1804.02934} {\path{arXiv:1804.02934}},
  \href {https://doi.org/10.1142/S0217751X18300119}
  {\path{doi:10.1142/S0217751X18300119}}.

\bibitem{Blackburn:2021rqm}
T.~G. Blackburn, A.~J. MacLeod, B.~King, {From local to nonlocal: higher
  fidelity simulations of photon emission in intense laser pulses}, New J.
  Phys. 23~(8) (2021) 085008.
\newblock \href {http://arxiv.org/abs/2103.06673} {\path{arXiv:2103.06673}},
  \href {https://doi.org/10.1088/1367-2630/ac1bf6}
  {\path{doi:10.1088/1367-2630/ac1bf6}}.

\bibitem{landau8}
E.~M. Lifshitz, L.~D. Landau, L.~P. Pitaevskii, {Electrodynamics of Continuous
  Media: Volume 8 (Course of Theoretical Physics)}, Butterworth-Heinemann,
  1984.

\bibitem{Heinzl:2009nd}
T.~Heinzl, D.~Seipt, B.~Kampfer, {Beam-Shape Effects in Nonlinear Compton and
  Thomson Scattering}, Phys. Rev. A 81 (2010) 022125.
\newblock \href {http://arxiv.org/abs/0911.1622} {\path{arXiv:0911.1622}},
  \href {https://doi.org/10.1103/PhysRevA.81.022125}
  {\path{doi:10.1103/PhysRevA.81.022125}}.

\bibitem{DiPiazza:2009zz}
A.~Di~Piazza, K.~Z. Hatsagortsyan, C.~H. Keitel, {Strong signatures of
  radiation reaction below the radiation dominated regime}, Phys. Rev. Lett.
  102 (2009) 254802.
\newblock \href {http://arxiv.org/abs/0810.1703} {\path{arXiv:0810.1703}},
  \href {https://doi.org/10.1103/PhysRevLett.102.254802}
  {\path{doi:10.1103/PhysRevLett.102.254802}}.

\bibitem{Heinzl:2013txd}
T.~Heinzl, C.~Harvey, A.~Ilderton, M.~Marklund, S.~S. Bulanov, S.~Rykovanov,
  C.~B. Schroeder, E.~Esarey, W.~P. Leemans, {Detecting radiation reaction at
  moderate laser intensities}, Phys. Rev. E 91~(2) (2015) 023207.
\newblock \href {http://arxiv.org/abs/1310.0352} {\path{arXiv:1310.0352}},
  \href {https://doi.org/10.1103/PhysRevE.91.023207}
  {\path{doi:10.1103/PhysRevE.91.023207}}.

\bibitem{Tang:2021qht}
S.~Tang, B.~King, {Pulse envelope effects in nonlinear Breit-Wheeler pair
  creation}, Phys. Rev. D 104~(9) (2021) 096019.
\newblock \href {http://arxiv.org/abs/2109.00555} {\path{arXiv:2109.00555}},
  \href {https://doi.org/10.1103/PhysRevD.104.096019}
  {\path{doi:10.1103/PhysRevD.104.096019}}.

\bibitem{Chen1998a}
S.~yuan Chen, A.~Maksimchuk, D.~Umstadter,
  \href{https://doi.org/10.1038%2F25303}{Experimental observation of
  relativistic nonlinear thomson scattering}, Nature 396~(6712) (1998)
  653--655.
\newblock \href {https://doi.org/10.1038/25303} {\path{doi:10.1038/25303}}.
\newline\urlprefix\url{https://doi.org/10.1038%2F25303}

\bibitem{khrennikov15}
K.~Khrennikov, J.~Wenz, A.~Buck, J.~Xu, M.~Heigoldt, L.~Veisz, S.~Karsch,
  \href{https://link.aps.org/doi/10.1103/PhysRevLett.114.195003}{Tunable
  all-optical quasimonochromatic thomson x-ray source in the nonlinear regime},
  Phys. Rev. Lett. 114 (2015) 195003.
\newblock \href {https://doi.org/10.1103/PhysRevLett.114.195003}
  {\path{doi:10.1103/PhysRevLett.114.195003}}.
\newline\urlprefix\url{https://link.aps.org/doi/10.1103/PhysRevLett.114.195003}

\bibitem{sakai15}
Y.~Sakai, I.~Pogorelsky, O.~Williams, F.~O'Shea, S.~Barber, I.~Gadjev,
  J.~Duris, P.~Musumeci, M.~Fedurin, A.~Korostyshevsky, B.~Malone, C.~Swinson,
  G.~Stenby, K.~Kusche, M.~Babzien, M.~Montemagno, P.~Jacob, Z.~Zhong,
  M.~Polyanskiy, V.~Yakimenko, J.~Rosenzweig,
  \href{https://link.aps.org/doi/10.1103/PhysRevSTAB.18.060702}{Observation of
  redshifting and harmonic radiation in inverse compton scattering}, Phys. Rev.
  ST Accel. Beams 18 (2015) 060702.
\newblock \href {https://doi.org/10.1103/PhysRevSTAB.18.060702}
  {\path{doi:10.1103/PhysRevSTAB.18.060702}}.
\newline\urlprefix\url{https://link.aps.org/doi/10.1103/PhysRevSTAB.18.060702}

\bibitem{Fleck:2020opg}
K.~Fleck, N.~Cavanagh, G.~Sarri, {Conceptual Design of a High-flux Multi-GeV
  Gamma-ray Spectrometer}, Sci. Rep. 10~(1) (2020) 9894.
\newblock \href {https://doi.org/10.1038/s41598-020-66832-x}
  {\path{doi:10.1038/s41598-020-66832-x}}.

\end{thebibliography}

\end{document}